# An Entertaining Example for the Usage of Bitwise Operations in Programming

**Hristina Kostadinova, Krasimir Yordzhev**
South-West University "Neofit Rilski"
Blagoevgrad, Bulgaria
kostadinova@swu.bg         yordzhev@swu.bg

**Abstract**: The present study is meant to fill in some information gaps occurring in the most widespread and well-known educational and reference literature about programming. The stress is laid on a very useful instrument - the bitwise operations, topic which is, unfortunately, seldom dealt with in most of the well-known books on programming. In addition, the research is very useful as regards the topic of overloading operators in any Object-oriented programming course. Given some appropriate examples, with the emphasis being laid on some particular and data structures language constructions, the results are quite interesting. The algorithm of solving the popular Sudoku puzzle is one such entertaining example.

## 1. INTRODUCTION

The use of bitwise operations is a popular and well-working method, used in C/C++ and Java programming languages. Unfortunately, in many widespread books on this topic there is only scarce or incomplete information as to how exactly bitwise operators function [4,14,15,16,20]. The aim of this article is to bridge this information gap to a certain extent and present a meaningful example of a programming task, where the use of bitwise operators is appropriate, in so far as to facilitate the work and increase the efficiency of the respective algorithm. Thus, we shall help the programming teachers in their work by presenting a good programming example, which demonstrates the use of bitwise operations in C/C++.

Some applications of the bitwise operations in the realization of algorithms designed to determine quantitative evaluations in the textile industry, as shown in [7,11,12].

Let $M = \{m_0, m_1, \ldots, m_{m-1}\}$, $|M| = m$ be a finite and discrete set. By the rule of thumb [17], each subset could be denoted by means of a Boolean vector $b(A) = \langle b_0, b_1, \ldots, b_{m-1} \rangle$, where $b_i = 1 \Leftrightarrow \mu_i \in A$ and $b_i = 0 \Leftrightarrow \mu_i \notin A$,

$i = 1,2,...,m-1$. A great memory economy could be achieved, if natural numbers are presented in a binary number system, where the number "zero" corresponds to null set, while the number $2^m - 1$, which in a binary number system is written by means of $m$ "ones", corresponds to the basic set $M$. Thus, a natural one to one correspondence between the integers of the closed interval $[0, 2^m - 1]$ and the set of all subsets of $M$ is achieved. The integer $a \in [0, 2^m - 1]$ corresponds to the set $A \subseteq M$, if for every $i = 1,2,...,m-1$ the $i$-bit of the binary representation of $a$ equals one, and if and only if $\mu_i \in A$. The need of the use of bitwise operations naturally arises in cases involving the computer realization of various operations with different sets of numbers.

Such an approach is also comfortable and quite efficient, given a relatively small cardinal number $m = |M|$ of the basic set $M$, and it also has a significant importance, in relation to the operating system and programming environment that is used. This is so, because $m$ bits are necessary to encode in the abovementioned method a set, which a subset of $M$, where $|M| = m$. And if k-bits are given for the type integer in the programming environment dealing with data, then $\left[\frac{m}{k}\right] + 1$ variables of that certain kind will be necessary, so as to put the abovementioned ideas into practice, where $[x]$ denotes the function "the whole part of $x$". Four bytes (thirty-two bits) are necessary to write a program that can solve a Sudoku puzzle in the size of $n^2 \times n^2$ if we use the set theory method at $n \leq 5$. In this case, every set of the kind $A = \{\alpha_1, \alpha_2, ..., \alpha_s\}$, where $1 \leq \alpha_i \leq n^2$, $i = 1,2,...,s$, $0 \leq s \leq n^2$, or the null set, could be simply encoded with an integer.

The programming realization of the problem of finding prime numbers using the method Sieve of Eratosthenes [1,5,8,19] has become a classical example for the use of sets in programming. Here we have to emphasize that a mistake has been made in [19], concerning this method's realization, and namely the number 1 has been reckoned among the set of the prime numbers. Such a mistake is inadmissible if the method is applied at schools, as the users may be confused.

How to construct a faster algorithm, solving the problem for receiving all $n \times n$ binary matrices, which contain exactly $k$ ones in each row and each column, with the help of the set theory and the operations over sets, is shown in [13].

Unfortunately the programming languages C/C++ and Java do not support a standard type "set", whereas the Pascal language does [1,5,8,19].

For this reason, if there should be a need to use the language of sets in the realization of some of our algorithms, and if there should be a need to write a program in some of these languages, we have to look for additional instruments to work with sets, such as, for example, the associative containers set and multiset, realized in Standard Template Library (STL) [15]**.** The template class set of the system of computer algebra "Symbolic C++" can also be used. The programming code is given in details in [9]. Of course, another class set can also be built, and specific methods of this class can be described, as a means of training. This is a good exercise, having in mind the fact that the cardinal number of the basic ("universal") set is not very big. For example the "standard" Sudoku puzzle has basic set the set of the integers from 1 to 9 plus the empty set.

We shall examine the entertaining problem with Sudoku solving algorithms, which is very interesting to many students. Sudoku is popular puzzle nowadays, populating the recreation pages of many newspapers and magazines, and it could also be found on many websites. Sudoku, or Su Doku, is a Japanese word (or phrase) meaning something like Number Place.

On the other hand, Sudoku matrices find an interesting combinatorial application. The connection between the set of all $m \times m$ permutation matrices (i.e. binary matrices which contain just one 1 in every row and every column) and the set of all $m \times m$ Sudoku matrices, is shown in [2].

## 2. TASK LAYOUT AND DESCRIPTION OF THE ALGORITHM USING THEORETIC SET OPERATIONS

Let $n$ be random positive integer and let $m = n^2$. Let $S = (s_{ij})$ be a square $m \times m$ matrix (square table), all elements in this table are integers, belonging to the closed interval $[1, m]$. The matrix $S$ is divided to $n^2$, $n \times n$ square submatrices, which are not intersected and will be called blocks, with the help of $n-1$ horizontal and $n-1$ vertical lines (the matrix $S$, when $n = 3$, is shown on fig.1).

|S₁₁|S₁₂|S₁₃|S₁₄|S₁₅|S₁₆|S₁₇|S₁₈|S₁₉|
|---|---|---|---|---|---|---|---|---|
|S₂₁|S₂₂|S₂₃|S₂₄|S₂₅|S₂₆|S₂₇|S₂₈|S₂₉|
|S₃₁|S₃₂|S₃₃|S₃₄|S₃₅|S₃₆|S₃₇|S₃₈|S₃₉|
|S₄₁|S₄₂|S₄₃|S₄₄|S₄₅|S₄₆|S₄₇|S₄₈|S₄₉|
|S₅₁|S₅₂|S₅₃|S₅₄|S₅₅|S₅₆|S₅₇|S₅₈|S₅₉|
|S₆₁|S₆₂|S₆₃|S₆₄|S₆₅|S₆₆|S₆₇|S₆₈|S₆₉|
|S₇₁|S₇₂|S₇₃|S₇₄|S₇₅|S₇₆|S₇₇|S₇₈|S₇₉|
|S₈₁|S₈₂|S₈₃|S₈₄|S₈₅|S₈₆|S₈₇|S₈₈|S₈₉|
|S₉₁|S₈₂|S₉₃|S₉₄|S₉₅|S₉₆|S₉₇|S₉₈|S₉₉|

Fig. 1:

Let denote by $A_{kl}, 1 \leq k, l \leq n$ the blocks in the above described *matrix* $S = (s_{ij})$. Then by definition if $s_{ij} \in A_{kl}$, then

$$(k-1)n < i \leq kn$$

and

$$(l-1) < j \leq (l-1).$$

Let $s_{ij}$ belong to the block $A_{kl}$ and let $i$ and $j$ be known. Then it is easy to guess that $k$ and $l$ can be calculated with the help of the formulas

$$k = \left[\frac{i-1}{n}\right] + 1$$

and

$$l = \left[\frac{j-1}{n}\right] + 1,$$

where, as usual, we denote by $[x]$ the function: whole part of the real number $x$.

We say that $S = (s_{ij})$, $1 \leq i, j \leq m = n^2$ is a *Sudoku matrix* if there is just one number of the set $Z_m = \{1, 2, \ldots, m = n^2\}$ in every row, every column and every block.

The Sudoku puzzle is quite popular these days. The user is given a Sudoku matrix, in which some of the elements have been erased. The missing elements could be equal to 0. The user's task is to restore the missing elements of the Sudoku matrix. It is supposed that the authors of the particular puzzle have chosen the missing elements in such a way that the problem has only one solution. This condition will be ignored. In this study we will build our programming product so that it can show the number of every possible solution. If the task has no solution this number will be zero.

The most popular Sudoku puzzles are with $n = 3$, i.e. $m = 9$.

We are going to describe an algorithm for creating a computer program, which can find all solutions (if there are any) of random Sudoku puzzle. For this purpose, we will use the knowledge of the set theory.

We examine the sets $R_i$, $C_j$ and $B_{kl}$, where $1 \leq i, j \leq m = n^2$, $1 \leq k, l \leq n$. For every $i = 1, 2, \ldots, m$, the set $R_j$ consists of all missing numbers in the $i$-th row of the matrix. Analogously we define the sets $C_j$, $j = 1, 2, \ldots, m$ correspondingly for the missing numbers in the $j$-th column and $B_{kl}$, $k, l = 1, 2, \ldots, n$ correspondingly for the missing numbers in the blocks $A_{kl}$ of $S$.

Whenever the algorithm comes into operation it traverses many times all of the elements $s_{ij} \in S$, such that $s_{ij} = 0$, i.e. these are the elements, which real values we have to find.

Let $s_{ij} = 0$ and let $s_{ij} \in A_{kl}$. We assume
$$P = R_i \cap C_j \cap B_{kl}.$$

Then the following three cases are possible:

i) $P = \phi$ (empty set). The task has no solution in this case;

ii) $P = \{d\}$, $d \in Z_m = \{1, 2, \ldots, m\}$, i.e. $|P|$: the number of the elements of $P$ is equal to 1 ($P$ is a set containing one element). Then the only one possibility for $s_{ij}$ is $s_{ij} = d$, i.e. we have found the unknown value of $s_{ij}$ in this case. After this we remove the common element $d$ from the sets $R_i$, $C_j$ and $B_{kl}$, and then we continue to the next zero element of the matrix $S$ (if there is such an element);

iii) $|P| \geq 2$. Then, we cannot say anything about the unknown value of $s_{ij}$ and we move on the next missing (zero) element of the matrix $S$.

We traverse all zero elements of the matrix $S$, until one the following events occur:

e1) For some $i, j \in \{1, 2, \ldots, m\}$ is true $s_{ij} = 0$, but $P = R_i \cap C_j \cap B_{kl} = \phi$;

e2) All elements in $S$ become positive;

e3) All zero elements of $S$ are traversed, but neither event e1, nor event e2 occur. In other words, for all the remaining zero elements in $S$, the above described case iii is always true.

In case any of the events e1 or e2 occurs, then the procedure is brought to a halt and the result visualizes on the screen.

In case event e3 occurs, then the algorithm has to continue operating by using other methods such as, for example, applying the "trial and error" method. In this particular case, the method would consist of the following steps:

We choose random $s_{ij} \in S$, such that $s_{ij} = 0$ and let $k = \left[\frac{i-1}{n}\right] + 1$, $l = \left[\frac{j-1}{n}\right] + 1$. Let $P = R_i \cap C_j \cap B_{kl} = \{d_1, d_2, \ldots, d_t\}$. Then for every $d_r \in P$, $r = 1, 2, \ldots, t$ we assume $s_{ij} = d_r$. Such an assumption is called a *random trial*. We count the number of all random trials, until the solution is found in the programming realization of the algorithm. Then, we solve the problem for finding the unknown elements of the Sudoku matrix, which contain one element less than the previous matrix. It is convenient to use a recursion here. The procedure is halted if event e1 or e2 occurs. It is absolutely certain to occur (i.e. there will not be "an infinite cycle"), because whenever we perform random trials, we reduce the number of the zero elements by 1.

## 3. BITWISE OPERATORS IN C/C++

Bitwise operations can be applied for integer data type only, i.e. they cannot be used for float and double types. For the definition of the bitwise operators in C/C++ and some of their elementary applications could be seen, for example in [3,6,10,18].

We assume as usual that bits numbering in variables starts from right to left, and that the number of the very right one is 0.

Let $x$, $y$ and $z$ are integer variables of one type, for which $w$ bits are needed. Let $x$ and $y$ are initialized and let the $z = x \alpha y$ assignment is made, where $\alpha$ is one of the operators & (bitwise AND), | (bitwise inclusive OR) or ^ (bitwise exclusive OR). For each $i = 0, 1, \ldots, w-1$ the new contents of the $i$ bit in $z$ will be as it is presented in the following table:

| The $i$ bit of $x$ | The $i$ bit of $y$ | The $i$ bit of $x$ & $y$ | The $i$ bit of $x \mid y$ | The $i$ bit of $x$^$y$ |
|---|---|---|---|---|
| 0 | 0 | 0 | 0 | 0 |
| 0 | 1 | 0 | 1 | 1 |
| 1 | 0 | 0 | 1 | 1 |
| 1 | 1 | 1 | 1 | 0 |

In case $z =\sim x$, if the $i$ bit of $x$ is 0, then the $i$ bit of $z$ becomes 1, and if the $i$ bit of $x$ is 1, then the $i$ bit of $z$ becomes 0, $i = 0,1,\ldots, w-1$

In case $k$ is a nonnegative integer, then the statement $z = x << k$; (bitwise shift left) will write in the $(i+k)$ bit of $z$ the value of the $k$ bit of $x$, where $i = 0,1,\ldots, w-k-1$, and the very right $k$ bits of $z$ will be filled by zeroes. This operation is equivalent to a multiplication of $x$ by $2^k$. The statement $z = x >> k$ functions similarly (bitwise shift right). But we must be careful here, as in various programming environments this operator has different interpretations – somewhere $k$ bits of $z$ from the very right place are compulsory filled by 0 (logical displacement), and elsewhere the very right $k$ bits of $z$ are filled with the value from the very left (sign) bit; i.e. if the number is negative, then the filling will be with 1 (arithmetic displacement). Therefore it is recommended that we use unsigned type of variables (if the opposite is not necessary) whenever we work with bitwise operations.

The effectiveness of the function computing $2^k$ is the direct result of the definition of the operator bitwise shift left, where $k$ is a nonnegative integer:

```
unsigned int Power2(unsigned int k) {
  return 1<<k;
}
```

To compute the value of the $i$ bit of an integer variable $x$ we can use the function:

```
int BitValue(int x, unsigned int i) {
  if ( (x & (1<<i) ) == 0 ) return 0;
    else return 1;
}
```

Bitwise operators are left associative.

The priority of operators in descending order is as follows: *bitwise complement* ~; *arithmetic operators* * (multiply), / (divide), % (remainder or modulus); *arithmetic operators* + (binary plus or add) - (binary minus or sub-tract); the *bitwise operators* << and >>; *relational operators* <, >, <=, >=, ==, !=; *bitwise operators* &,^ and |; *logical operators* && and ||.

## 4. DESCRIPTION OF THE CLASS SET BY MEANS OF OVERLOADING OPERATORS AND USING BITWISE OPERATIONS

```
class Set_N {
int n;
public:
 /*Constructor without parameter – creates empty set*/
 Set_N();
 /*Constructor with parameter – creates the set, corresponding to every bit position of the parameter*/
 Set_N(int k);
/*Returns the integer encoding of the set*/
 int get_n() const;
/*Overloading operators *, +, -, ==, <=*/
 Set_N operator * (Set_N const &);
 Set_N operator + (Set_N const &);
 Set_N operator + (int k);
 Set_N operator - (Set_N const &);
 Set_N operator - (int k);
 bool operator == (Set_N const &);
 bool operator <= (Set_N const &);
/*Checks if the first parameter belongs to the given set*/
 bool in(int k);
/*Destructor*/
 ~Set_N();
};
```

The realization of the Set_N methods is described below:

```
Set_N::Set_N() {n = 0;}
Set_N::Set_N(int k) {n = k;}
int Set_N::get_n() const {
 return n;
}
Set_N Set_N::operator * (Set_N const & s) {
 return (this->n) & s.get_n();
}
Set_N Set_N::operator + (Set_N const & s) {
 return (this->n) | s.get_n();
}
Set_N Set_N::operator + (int k) {
 return (this->n) | (1<<(k-1));
```

```
}
Set_N Set_N::operator - (Set_N const & s) {
 int temp = 0;
 temp = this->n ^ s.get_n();
 return (this->n) & temp;
}
Set_N Set_N::operator - (int k) {
 return (this->n) ^ (1<<(k-1));
}
bool Set_N::operator == (Set_N const & s) {
 return ((this->n ^ s.get_n()) == 0);
}
bool Set_N::operator <= (Set_N const & s) {
 return ((this->n & s.get_n()) == n);
}
bool Set_N::in(int k, Set_N const & s) {
 return this->n & (1<<(k-1));
}
```

## 5. REFERENCES

[1] Dahl, D. J., Dijkstra, E. W., Hoare, C. A. R. (1972) *Structured Programming.* Academic Press Inc.
[2] Dahl, G. (2009) Permutation matrices related to Sudoku. *Linear Algebra and its Applications* 430, 2457–2463.
[3] Davis, S. R. (2000) *C++ for dummies.* IDG Books Worldwide.
[4] Horstmann, C. S. (1999) *Computing concepts with C++ essentials*. John Wiley & Sons.
[5] Jensen, K., Wirth, N. (1985) *Pascal User Manual and Report*. 3rd ed., Springer-Verlag.
[6] Kernigan, B. W., Ritchie D. M. (1998) *The C programming Language*. AT&T Bell Laboratories.
[7] Kostadinova, H., Yordzhev, K. (2010) A Representation of Binary Matrices. *Mathematics and education in mathematics,* v. 39, 198-206.
[8] Price, D. (1983) *UCSD Pascal A Considerate Approach*. Prentice-Hall.
[9] Shi, T. K., Steeb, W.-H., Hardy, Y. (2000)*Symbolic C++: An Introduction to Computer Algebra using Object-Oriented Programming*. Springer.

[10] Schildt, H. *Java 2 A Beginner's Guide.* (2001) McGraw-Hill.
[11] Yordzhev, K. (2009) An example for the use of bitwise operations in programming. *Mathematics and education in mathematics,* v. 38, 196-202
[12] Yordzhev, K., Kostadinova, H. (*2010)* Mathematcal Modeling of the Weaving Structure Design *Mathematics and education in mathematics,* v. 39, 212-220.
[13] Yordzhev, K., Stefanov, S. (2003) On Some Applications of the Consept of Set in Computer Science Course. *Mathematics and Educations in Mathematics*, v.32, 249-252.
[14] Азълов П. (2005) *Информатика Езикът C++ в примери и задачи за 9-10 клас.* София, Просвета.
[15] Азълов П. (2008) *Обектно-ориентирано програмиране – Структури от данни и STL*. София, Сиела.
[16] Крушков, Х. (2006) *Практическо ръководство по програмиране на C++.* Пловдив, Макрос.
[17] Наков, П., Добриков П. (2005) Програмиране =++ Алгоритми, Трето издание, София, ISBN 954-8905-16-X.
[18] Романов, Е. Л. (2004) *Практикум по программированию на C++*. Петербург, БХВ.
[19] Стоянова, Р., Гочев, Г. (1994) *Програмиране на Pascal с 99 примерни програми*. София, Paraflow.
[20] Тодорова М. (2002) Програмиране на C++. Част I, Част II, София, Сиела.